\documentclass[letterpaper,12pt,abstracton,smallheadings,tbtags]{scrartcl}

\usepackage[left=1.25in,right=1in,top=0.75in,bottom=0.75in]{geometry}

\setkomafont{descriptionlabel}{\upshape\bfseries}
\addtokomafont{captionlabel}{\sffamily\bfseries\small}
\setkomafont{caption}{\sffamily\small}
\setbibpreamble{\small\vspace{-3em}}

\newcommand*{\appendixmore}{%
\renewcommand*{\othersectionlevelsformat}[1]{%
\ifthenelse{\equal{##1}{section}}{\appendixname~}{}%
\csname the##1\endcsname\autodot\enskip}
\renewcommand*{\sectionmarkformat}{%
\appendixname~\thesection\autodot\enskip}
}

 \usepackage[T1]{fontenc}
\usepackage[latin1]{inputenc}

\usepackage{graphicx}

\usepackage{amsmath,nccmath,braket,amssymb}
\usepackage{amsthm}

\usepackage{dsfont,mathrsfs}

\newtheorem{problem}{Problem}
\newtheorem{theorem}{Theorem}
\newtheorem{definition}{Definition}
\newtheorem{lemma}{Lemma}

\newcommand\abs[1]{\left|#1\right|}

\newcommand\E[1]{\mathds E\Set{#1}}

\newcommand\PM[1][t-1]{P_{M_{#1}}}
\newcommand\tPM[1][t-1]{\widetilde P_{M_{#1}}}

\newcommand\PYM[1][t]{P_{Y_{#1},M_{#1-1}}}
\newcommand\PYMempty{P_{Y,M}}
\newcommand\tPYM[1][t]{\widetilde P_{Y_{#1},M_{#1-1}}}
\newcommand\tPYMempty{\widetilde P_{Y,M}}

{\catcode`\|=\active \gdef\conditional#1{\begingroup 
\mathcode`\|32768\let|\BraVert\left({#1}\right)\endgroup}
}

\let\PPr\Pr
\renewcommand\Pr[1]{\PPr\conditional{#1}}

\newcommand*\defined{\triangleq}

\newcommand*\sC{\mathscr C}
\newcommand*\sG{\mathscr G}
\newcommand*\sL{\mathscr L}
\renewcommand*\P[1]{\mathcal P^{#1}}
\newcommand*\J{\mathcal J}
\newcommand*\M{\mathcal M}

\newcommand*\V{\mathcal V}
\newcommand*\W{\mathcal W}
\newcommand*\X{\mathcal X}
\newcommand*\Y{\mathcal Y}
\newcommand*\Z{\mathcal Z}

\begin{document}
\title{\large A Decision Theoretic Framework for Real-Time Communication}
\author{\normalsize Aditya Mahajan
\and \normalsize Demosthenis Teneketzis
}
\date{\small Department of EECS,
University of Michigan,
Ann Arbor, MI -- 48109-2122, USA.\\
\texttt{\{adityam,teneket\}@eecs.umich.edu}}
\maketitle
\thispagestyle{empty}

\vspace{-3em}
\begin{abstract}
  \noindent We consider a communication system in which the
  outputs of a Markov source are encoded and decoded
  in \emph{real-time} by a finite memory receiver, and the distortion measure 
  does not tolerate delays. The objective is to choose designs, i.e.  
  real-time encoding, decoding and memory update strategies that minimize a 
  total expected distortion measure. This is a dynamic team problem with 
  non-classical information structure~\cite{Witsenhausen:1971}. We use the 
  structural results of~\cite{Teneketzis:2004} to develop a sequential 
  decomposition for the finite and infinite horizon problems. Thus, we obtain 
  a systematic methodology for the determination of jointly optimal encoding 
  decoding and memory update strategies for real-time point-to-point 
  communication systems.
\end{abstract}

  \noindent\emph{Keywords:} Real-time communication, finite-delay 
  communication, zero-delay communication, joint source-channel coding, 
  Markov decision theory

\vspace{-1em}
\section{Introduction}\label{sec:intro}
\vspace{-1em}

Real-time communication problems arise in controlled   
decentralized systems where information must be exchanged between various 
nodes of the system and decisions based on the communicated information must  
be made in real-time. Such systems include QoS (delay) requirements and 
distributed routing in wired, wireless and sensor networks, traffic flow 
control in transportation networks, resource allocation and consensus in 
partially synchronous systems and decentralized resource allocation problems 
in economic systems.

We consider point-to-point real-time communication system as shown in 
Figure~\ref{fig:system}, which is the simplest system of this class. A better 
understanding of this case is needed before  generalizing to 
multi-terminal systems.  In the system under consideration, the outputs of a 
Markov source are  encoded in \emph{real-time} into a sequence of random 
variables.  This sequence is transmitted through a discrete memoryless 
channel (DMC) to a receiver with \emph{finite memory}. At each time instant 
$t$, using the current channel output and its current memory content,
the receiver updates its memory and estimates the source output at $t$. The 
system designer has to choose real-time encoding, decoding and memory update 
rules that minimize an expected total distortion.

Real-time or finite-delay communication problems have been considered in the 
past. For an extensive literature survey we refer the reader 
to~\cite{Teneketzis:2004,MahajanTeneketzis:2005}. Here we will only refer to 
the papers most relevant to our philosophy and approach.  Real-time encoding 
and decoding with limited memory make the standard Information theoretic 
techniques inappropriate for this problem. Most of the results of Information 
theory are based on some form of the law of large numbers, which becomes 
applicable only when we consider sufficiently long sequences. This problem 
does not have enough structure to use encoding and decoding of typical 
sequences.  Hence, we consider a decision theoretic approach to the problem.

Decision theoretic approaches to real-time communication similar in spirit to 
ours have appeared in~\cite{Witsenhausen:1978,BorkarMitterTatikonda:2001, 
WalrandVaraiya:1983,LipsterShiryayen:II,Teneketzis:2004}.  Real-time 
communication problems for noiseless channel were studied 
in~\cite{Witsenhausen:1978,BorkarMitterTatikonda:2001}.
Real-time encoding decoding problems for a noisy channel and noiseless 
feedback were studied in~\cite{WalrandVaraiya:1983,LipsterShiryayen:II}.  
These problems share a common feature that at every stage
the encoder has perfect knowledge of the information available to the
decoder/receiver. The case of a noisy channel and no feedback does not share
this feature. Real-time communication through noisy
channels and no feedback was investigated in~\cite{Teneketzis:2004} and 
structural results for optimal real-time encoding and decoding strategies 
were obtained.  However, to the best of our knowledge, the problem of 
obtaining jointly optimal real-time encoders, decoders and memory update 
rules has not been considered by anyone so far. In this paper we present a 
methodology for this joint optimization. We present the key ideas and 
fundamental results here, and refer the reader 
to~\cite{MahajanTeneketzis:2005} for details and extensions. 

The remainder of the paper is organized as follows. In Section~\ref{sec:prob} 
we formally define the problem, in Section~\ref{sec:structural} we restate 
the structural results of~\cite{Teneketzis:2004}, in Section~\ref{sec:joint} 
we present the joint optimization of encoder, decoder and memory update. In 
Sections~\ref{sec:finite} and~\ref{sec:infinite} we consider the finite and 
infinite horizon time homogeneous cases. We discuss some salient  points in 
Section~\ref{sec:discussion} and conclude in Section~\ref{sec:conclusion}.

\textbf{Notation:} When using English letters to represent a
variable, we use the standard notation of using uppercase letters
($X,Y,Z$) for denoting random variables and lowercase letters for
denoting their realization ($x,y,z$). While representing a function of
random variables as a random variable ($\PM[],\PYMempty$), we use a
tilde above the variable to denote its realization
($\tPM[],\tPYMempty$). When using Greek letters to represent a random
variable ($\pi,\varphi$), we use a tilde above the variable to denote
its representation ($\tilde\pi, \tilde\varphi$). We also use the standard 
short-hand notation of $x_s^t$ to represent the sequence $x_s,\dots,x_t$
, $x_1^t$ is abbreviated to $x^t$ and similar notation for random variables 
and functions. 

\vspace{-1em}
\section{Problem Formulation}\label{sec:prob}
\vspace{-1em}
\begin{figure}[tbp]
  \centering
  \includegraphics{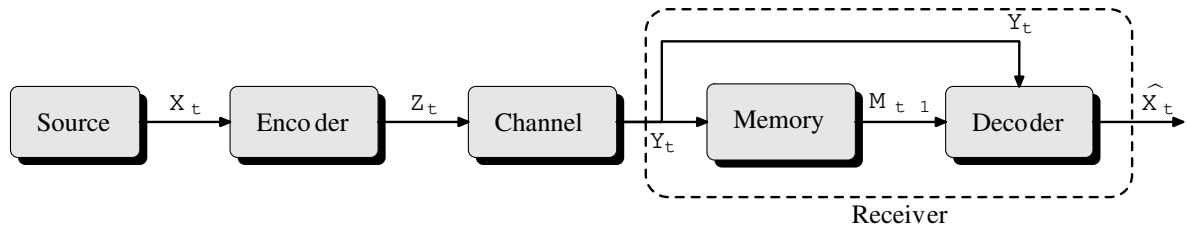}
  \caption{Real-Time Communication System}
  \label{fig:system}
\end{figure}
We now give a formal description of the problem under consideration.  
Consider a discrete time communication system shown in 
Figure~\ref{fig:system}. A first order Markov source produces a random 
sequence $X_1,\dots,X_T$.  For simplicity of exposition we assume that $X_t$ 
belongs to a finite alphabet $\X \defined \set{1,\dots,\abs\X}$.

At each stage $t$, the encoder can transmit a symbol $Z_t$ taking values in a 
finite alphabet $\Z \defined \set{1,\dots,\abs\Z}$. This encoded symbol must 
be generated in real-time, i.e.,
\begin{equation}
  Z_t = c_t(X_1,\dots,X_t),\qquad t = 1,\dots,T,
\end{equation}
and transmitted through a $\abs\Z$-input $\abs\Y$-output discrete memoryless 
channel producing the sequence $\set{Y_1,\dots,Y_T}$, with each $Y_t$ 
belonging to an alphabet $\Y \defined \set{1,\dots,\abs\Y}$. The transition 
probabilities of the channel is given by
\begin{equation}\label{eq:channel}
  \Pr{y_t | x^t, z^t, y^{t-1}} = \PPr\conditional{y_t|z_t} = P_t(y_t,z_t).
\end{equation}
At the receiver, the most that could be accessible at stage $t$ is the 
subsequence $Y_1,\dots,Y_t$. However, we assume that the receiver has a 
memory of $\log_2\abs\M$ bits. So, after some time, all the past observations 
can not be stored and the receiver must selectively \emph{shed} information.
We model this by assuming that the contents of the memory belong to a finite 
alphabet $\M \defined \set{1,\dots,\abs\M}$. The memory is arbitrarily 
initialized with $M_0 = 1$ and then updated at each stage according to the 
rule
\begin{equation}\label{eq:memory}
  M_t = l_t(Y_t,M_{t-1}), \qquad t = 1,\dots,T-1.
\end{equation}

The objective of the decoder is to generate an estimate of the source output 
in real-time. This estimate $\widehat X_t$ has to be generated from the 
present channel output $Y_t$ and the memory contents $M_{t-1}$, by some 
decoding rule, i.e.,
\begin{equation}
  \widehat X_t = g_t(Y_t, M_{t-1}),\qquad t = 1,\dots,T.
\end{equation}

The performance of the system is defined by way of a sequence of distortion 
functions. For each $t$,
\(
  \rho_t : \X\times\X\to [0, \infty).
\)
is given. Then, $\rho_t(X_t,\widehat X_t)$ measures the distortion at
stage $t$.

A choice $(c,g,l)$ of decision rules for all stages is called a 
\emph{design}, where $c \defined (c_1,\dots,c_T)$, $g \defined 
(g_1,\dots,g_T)$ and $l \defined (l_1,\dots,l_{T-1})$. The performance of a 
design is quantified by the expected distortion under that design, which is 
given by
\begin{equation}\label{eq:performance}
  \J(c,g,l) \defined \E{\medop\sum_{t=1}^{T}\rho_t(X_t,\widehat X_t )| c,g,l}.
\end{equation}
The optimization problem that we consider is as follows:
\begin{problem}\label{prob:P1}
  Assume that the encoder and the receiver know the statistics of the 
  source (i.e. PMF of $X_1$ and the transition probabilities 
  $P_{X_{t+1}|X_t}$), the channel transition matrix $P_t$, the distortion 
  function $\rho_t(\cdot,\cdot)$ and a time horizon $T$. \emph{Choose} a 
  design $(c^*,g^*,l^*)$ that is optimal with respect to the performance 
  criterion of~\eqref{eq:performance}, i.e.,
  \begin{equation}
    \J(c^*,g^*,l^*) = \J^* \defined \min_{\substack{c \in \sC^T \\ g \in 
    \sG^T \\ l \in \sL^{T-1}}} \J(c,g,l),
  \end{equation}
  where $\sC^T \defined \sC_1\times\dots\times\sC_T$, where $\sC_t$ is
  the family of functions from $\X^t \to \Z$, $\sG^T \defined
  \sG\times\dots\times\sG$ ($T$-times), where $\sG$ is the family of
  functions from $\Y\times\M\to\X$ and $\sL^{T-1} \defined
  \sL\times\dots\times\sL$ ($(T-1)$-times), where $\sL$ is the family
  of functions from $\Y\times\M\to\M$.
\end{problem}

The problem belongs to the class of decentralized \emph{dynamic team} 
problems with non-classical information structure. Such problems are 
difficult to solve as they are non-convex functional optimization problems.
We can view the problem as a sequential stochastic optimization 
problem~\cite{Witsenhausen:1973,Witsenhausen:1975c} by a fictitious 
partitioning of stage $t$ into three parts. The encoder transmits at $t^+$, 
the decoder makes a decision at $(t+\frac12)$ and the memory is updated at 
$(t+1)^-$. Now we have a stochastic optimization problem with a horizon of 
$3T$ where the decision makers can be ordered in advance, thus the problem is 
sequential.  Witsenhausen~\cite{Witsenhausen:1973} presented a general 
framework to work with sequential stochastic optimization problems by 
converting them into \emph{standard form}. The solution methodology presented 
therein is applicable only to finite horizon problems and can not be extended 
to infinite horizon problems. We exploit the  structural
results of~\cite{Teneketzis:2004} to obtain a solution methodology which can 
be extended to infinite horizon problems.  For completeness of presentation 
we summarize the structural results of~\cite{Teneketzis:2004} next.

\vspace{-1em}
\section{Structural Results}\label{sec:structural}
\vspace{-1em}
\begin{definition}
  Let $\PM[t]$ be the encoder's belief about the memory contents of the 
  receiver, i.e.,
  \begin{equation}
    \PM[t](m) \defined \Pr{M_t = m | X^t, Z^t, c^{t}, l^{t}}.
  \end{equation}
\end{definition}
For a particular realization $x^t$ and an arbitrary (but fixed)  choice of 
$c^{t}$, $l^{t}$, the realization of $\PM[t]$ denoted by $\tPM[t]$, is a PMF 
on $M_t$ and belongs to $\P\M$, the space of PMFs on $\M$. If $X^t$ is random 
vector and $c^{t}$, $l^{t}$ are arbitrary (but fixed) functions, then 
$\PM[t]$ is a random vector belonging to $\P\M$.

\begin{theorem}[Structure of Optimal Encoder]\label{thm:opt-enc}
  Consider the problem of minimizing the expected distortion given 
  by~\eqref{eq:performance} for any arbitrary (but fixed) decoder $g$ and 
  memory update $l$. Then, without loss in optimality, one can restrict 
  attention to encoding rules of the form
  \begin{equation}
    Z_t = c_t(X_t, \PM[t-1]), \qquad t = 2,\dots,T.
  \end{equation}
\end{theorem}

\begin{theorem}[Structure of Optimal Decoder]\label{thm:opt-dec}
  Consider the problem of minimizing the expected distortion given 
  by~\eqref{eq:performance} for any arbitrary (but fixed) encoder $c$ and 
  memory update rule $l$. Then, the design of an optimal decoder is a 
  filtering problem and an optimal decoding rule $g^*$ is given by
  \begin{align}
    \hat x_t &= g_t^*(y_t,m_{t-1}) = \tau_t\big(\xi_t(y_t,m_{t-1})\big), \\
    \intertext{where}
    \xi_t(y,m)(x) &= \Pr{X_t = x | Y_t = y, M_{t-1} = m }, \\
    \intertext{and}
    \tau_t(\xi_t) &= \arg \min_{\hat x} \medop\sum_{x \in \X} \rho_t(x,\hat 
    x)\xi_t(x).
  \end{align}
\end{theorem}

See~\cite{Teneketzis:2004} for a proof of these theorems.

\vspace{-1em}
\subsection{Implication of Structural Results}
The structural results simplify the problem as follows:
\begin{enumerate}
\item Theorem~\ref{thm:opt-enc} implies that at each stage $t$,
  without loss in optimality, we can restrict attention to encoders
  belonging to $\sC_S$, the family of functions from $\X\times\P\M$ to
  $\Z$. Thus, at each stage, we can restrict to optimizing over a
  fixed (rather than time-varying) domain.
\item Theorem~\ref{thm:opt-dec} implies that the structure of an optimal 
  decoders is a deterministic function of $\rho_t$, the distortion measure at 
  time $t$ and $\xi_t$, the conditional PMF at time $t$, which in turn 
  depends only on the choice of decision rules $c^t$ and $l^{t-1}$.  Thus, an 
  optimal decoder at time $t$ can be written as $g^*_t = g^*_t(c^t,l^{t-1})$, 
  implying that an optimal decoder obtained by Theorem~\ref{thm:opt-dec} can 
  be
  expressed in terms of the encoder and memory update rule as $g^*(c,l)$. For 
  any design define
  \begin{equation}
    \widetilde\J(c,l) \defined 
    \J\big(c,g^*(c,l),l\big),\label{eq:perf-opt-dec}
  \end{equation}
  and consider the following problem:
  \begin{problem}\label{prob:P2}
    Under the assumptions of Problem~\ref{prob:P1}, choose a design
    $(c^*,l^*)$ that is optimal with respect to the performance
    criterion of~\eqref{eq:perf-opt-dec}, i.e.,
    \begin{equation}
      \widetilde\J(c^*,l^*) = \widetilde\J^* = \inf_{\substack{c^* \in 
      \sC_S^T \\ l \in \sL^T}} \J\big(c,g^*(c,l),l\big),
    \end{equation}
    where $\sC_S^T \defined \sC_S\times\dots\times\sC_S$ ($T$-times)
  \end{problem}
  Clearly,
  \begin{math}
    \widetilde\J^* = \J^*
  \end{math}
  i.e., the design $(c^*,g^*,l^*)$ obtained by an optimal solution
  $(c^*,l^*)$ of Problem~\ref{prob:P2}, along with an optimal decoder
  $g^*(c^*,l^*)$ obtained by Theorem~\ref{thm:opt-dec}, is an optimal
  solution for Problem~\ref{prob:P1}.
\end{enumerate}

In the next section we provide a sequential decomposition for
Problem~\ref{prob:P2}.

\vspace{-1em}
\section{Joint Optimization}\label{sec:joint}
\vspace{-1em}

The critical step in obtaining an optimization methodology based on 
sequential decomposition is identifying an information state sufficient for 
performance evaluation of the system.  In this section, we give expressions 
for an information state and explain how to obtain a sequential decomposition 
of the problem.  The intuition behind our approach is as follows. The 
distortion at stage $t$ depends on $X_t$ and $\widehat X_t$. We need to find 
a \emph{field basis} and \emph{conditional basis} for $\widehat X_t$ 
(see~\cite{Witsenhausen:1971}) for each agent at each stage. However, just 
finding a field and conditional bases is not sufficient. These combined bases 
must form a \emph{state} (in the sense of~\cite{Witsenhausen:1976}) for the 
purpose of performance evaluation. Suppose $\pi_t$ and $\varphi_t$ are the 
information states of the encoder and memory update respectively.  They need 
to satisfy the following properties:
\begin{enumerate}
\item $\pi_t$ is a function only of the encoder's information and the past 
  encoding and memory update rules. Any choice of the  present encoding rule 
  $c_t$ together with $\pi_t$ determine $\varphi_t$, the information state 
  for the memory update at the next step.
\item $\varphi_t$ is a function only of the receiver's information and the 
  past encoding and memory update rules. Any choice of the present memory 
  update rule $l_t$ together with $\varphi_t$ determine $\pi_{t+1}$, the 
  information state for the encoder at the next step.
\item At each stage both the encoder and the receiver can evaluate the 
  expected cost to go from their respective information state and choice of 
  present and future decision rules.  \emph{This expectation is conditionally 
  independent of the past decision rules, conditioned on the current 
  information state}.
\end{enumerate}
The above properties can be written more formally as follows:
\begin{description}
  \item[(S1a)] $\pi_t$ is a function of $x^t$, $c^{t-1}$ and $l^{t-1}$.
  \item[(S1b)] $\varphi_t$ is a function of $y_t$, $m_{t-1}$, $c^t$ and 
    $l^{t-1}$.
  \item[(S2a)] $\varphi_t$ can be determined from $\pi_t$ and $c_t$.
  \item[(S2b)] $\pi_{t+1}$ can be determined from $\varphi_t$ and $l_t$.
  \item[(S3)] For the purpose of performance evaluation, $\pi_t$ 
    \emph{absorbs} the effect of $c^{t-1},l^{t-1}$ and $\varphi_t$ 
    \emph{absorbs} the effect of $c^t,l^{t-1}$ on expected future distortion, 
    i.e.\\[1.2em]
  $ \E{\sum\limits_{s=t}^T \rho_s(X_s,\widehat X_s) |
    c,g,l} =
  \E{\sum\limits_{s=t}^T \rho_s(X_s,\widehat X_s) | \pi_t, c_t^T, l_t^T}$ \\
  $\hphantom{ \E{\sum\limits_{t=s}^T \rho_t(X_t,\widehat X_t) |
      c,g,l}} = \E{\sum\limits_{s=t}^T\rho_s(X_s,\widehat X_s) |
      \varphi_t, c_{t+1}^T, l_t^T}$,\\
      or alerntively
    \item[(S3$^\star$)]
 $\E{\rho_t(X_t,\widehat X_t) | c,g,l} = \E{\rho_t(X_t, \widehat X_t) | 
 \pi_t, c_t}$ \\
  $\hphantom{ \E{\rho_t(X_t,\widehat X_t) | c,g,l}} = \E{\rho_t(X_t, \widehat 
  X_t) | \varphi_t, l_t}.$
\end{description}
Properties \textbf{(S1)},\textbf{(S2)},\textbf{(S3)} are equivalent to 
properties \textbf{(S1)},\textbf{(S2)},\textbf{(S3$^\star$)}. \textbf{(S1)} 
and \textbf{(S2)} imply that $\pi_t$ and $\varphi_t$ are \emph{states} and 
\textbf{(S3)} ensures that $\pi_t$ and $\varphi_t$ absorb the effect of past 
decision rules on expected future distortion.  Thus, they are sufficient for 
the purpose of performance evaluation. In this section we find information 
states $\pi_t$ and $\varphi_t$ that satisfy \textbf{(S1)}--\textbf{(S3)}. We 
define the following:
\begin{definition}
  Let $\PYM[t]$ be the encoder's belief about the channel output and memory 
  contents of the receiver, i.e.,
  \begin{equation}
    \PYM[t](y,m)\defined  \Pr{Y_t = y, M_{t-1} = m | X^t, Z^t, c^t, l^{t-1}}.
  \end{equation}
\end{definition}
For a particular realization $x^t$ and a particular choice $c^{t}$,
$l^{t-1}$, the realization of $\PYM[t]$, denoted by $\tPYM[t]$, is a PMF on
$(Y_t,M_{t-1})$ and belongs to $\P{\Y\times\M}$, the space of PMFs on
$\Y\times\M$. If $X^t$ is a random vector and $c^t$, $l^{t-1}$ are
arbitrary (but fixed) functions, then $\PYM[t]$ is a random vector belonging 
to $\P{\Y\times\M}$.

\begin{lemma}
  \label{lemma:updates}
  At each stage $t$,
  \begin{enumerate}
    \item there is a deterministic function $\nu_t(\cdot)$ such that
      \(
        \PYM[t] = \nu_t(\PM[t-1],Z_t),
      \)
    \item there is a deterministic function $\psi_t(\cdot)$ such that
      \(
        \PM[t] = \psi_t(\PYM[t],l_t).
      \)
  \end{enumerate}
\end{lemma}
\begin{proof}
  See~\cite{MahajanTeneketzis:2005}.
\end{proof}

\begin{definition}
  Let $\Pi$ (resp\@. $\Phi$) be the space of probability measures on 
  $\X\times\P\M$ (resp\@. $\X\times\P{\Y\times\M}$). Define $\pi_t$ and 
  $\varphi_t$ as follows:
  \begin{align}
    \pi_t &= \Pr{X_t,\PM[t-1]}, \\
    \varphi_t &= \Pr{X_t, \PYM[t]},
  \end{align}
  where $\pi_t$ (resp\@. $\varphi_t$) belongs to $\Pi$ (resp\@. $\Phi$).
\end{definition}

\begin{theorem}\label{thm:info-state}
  $\pi_t$ and $\varphi_t$ are the information states for the encoder and 
  memory update respectively, i.e.,
  \begin{enumerate}
    \item there is a linear transformation $Q_t(c_t)$ such that
      \begin{equation}\label{eq:phi_update}
        \varphi_t = Q_t(c_t)\pi_t,
      \end{equation}
    \item there is a linear transformation $\widehat Q_t(l_t)$ such that
      \begin{equation}\label{eq:pi_update}
        \pi_{t+1} = \widehat Q_t(l_t)\varphi_t,
      \end{equation}
    \item for any choice of $c$ and $l$, the expected conditional 
      instantaneous cost can be expressed as
      \begin{equation}\label{eq:cost_state}
        \E{\rho_t(X_t,\widehat X_t) | c^t, g_t^*(c^t,l^{t-1}), l^{t-1}} = 
        \tilde\rho_t(\varphi_t).
      \end{equation}
      where $g^*_t(c^t,l^{t-1})$ is an optimal decoding rule corresponding to 
      $c^t$, $l^{t-1}$ and $\tilde\rho_t(\cdot)$ is a deterministic function.
  \end{enumerate}
\end{theorem}
\begin{proof}
  This follows from Lemma~\ref{lemma:updates}.  
  See~\cite{MahajanTeneketzis:2005} for detailed proof.
\end{proof}

The choice of functions $c^t$, $l^{t-1}$, $g^*_t(c^t,l^{t-1})$ make the 
variable $\widehat X_t$ a random variable with  well defined distribution.  
Thus, the performance criterion of~\eqref{eq:perf-opt-dec} can be rewritten as
\begin{equation}\label{eq:eq_cost}
  \begin{split}
  \E{\sum_{t=1}^{T}\rho_t(X_t, \widehat X_t) | c, g^*(c,l), l} &= 
  \sum_{t=1}^{T}\E{\rho_t(X_t, \widehat X_t) | c^t, g_t^*(c^t,l^{t-1}), 
  l^{t-1}} \\
  &= \medop\sum_{t=1}^T \tilde\rho_t(\varphi_t).
  \end{split}
\end{equation}
Notice that~\eqref{eq:phi_update} and~\eqref{eq:pi_update} imply that $\pi_t$ 
and $\varphi_t$ are states, i.e. they satisfy \textbf{(S1)} and 
\textbf{(S2)}. Moreover,~\eqref{eq:cost_state} and~\eqref{eq:eq_cost} imply 
that $\pi_t$ and $\varphi_t$ are sufficient for performance evaluation, i.e., 
satisfy \textbf{(S3)}. Hence Theorem~\ref{thm:info-state} implies that 
Problem~\ref{prob:P2} is equivalent to the following deterministic 
optimization problem:
\begin{problem}\label{prob:P3}
  Consider a deterministic system which evolves as follows:
  \begin{alignat}{2}
    \varphi_t &= Q_t(c_t)\pi_t, &\qquad t &= 1,\dots,T,\\
    \pi_{t+1} &= \widehat Q_t(l_t)\varphi_t, &\qquad t &= 1,\dots,T-1,
  \end{alignat}
  where $c_t$ and $l_t$ are functions belonging to $\sC_s$ and $\sL$
  respectively and $Q_t(\cdot)$ and $\widehat Q_t(\cdot)$ are deterministic
  transforms depending on $c_t$ and $l_t$ respectively. The initial state
  $\pi_1$ of the system is known. If the system is in state $\varphi$ at 
  stage $t$, it incurs a cost $\tilde\rho_t(\varphi)$.

  The optimization problem is to obtain decision rules $c \defined 
  (c_1,\dots,c_T)$, $l \defined (l_1,\dots,l_{T-1})$ to minimize the total 
  cost over horizon $T$, i.e., find optimal design $(c^*,l^*)$ such that
  \begin{equation}
    \widetilde\J(c^*,l^*) = \widetilde\J^* = \inf_{\substack{c^* \in \sC_S^T 
    \\ l^* \in \sL^T}} \medop\sum_{t=1}^T \tilde\rho_t(\varphi_t).
  \end{equation}
\end{problem}

This is a classical deterministic control problem;  optimal functions 
$(c^*,l^*)$ are determined by the nested optimality equations given below.

\begin{theorem}\label{thm:DP}
  An optimal design $(c^*,l^*)$ for Problem~\ref{prob:P3} (and consequently 
  for Problem~\ref{prob:P2} and thereby for Problem~\ref{prob:P1}) can be 
  determined by the solution of the following nested optimality equations:
  \begin{align}
    \widehat V_T(\varphi) &\equiv 0, \\
    V_t(\pi_t) &= \inf_{c_t \in \sC_S } \Big[ 
    \tilde{\rho}_t\big(Q_t(c_t)\pi\big) + \widehat V_t\big(Q_t(c_t)\pi\big) 
    \Big], & t &= 1,\dots,T, \\
    \widehat V_t(\varphi) &= \min_{l_t \in \sL} \Big[ V_{t+1}\big(\widehat 
    Q_t(l_t)\varphi\big) \Big], & t &= 1,\dots,T-1.
  \end{align}
\end{theorem}
\begin{proof}
  This is a standard result, see~\cite[Chapter~2]{Whittle:1982}
\end{proof}

\vspace{-1em}
\section{Time Homogeneous System --- Finite Horizon Case}
\label{sec:finite}
\vspace{-1em}

For many applications the system is time-homogeneous, that is, the source is 
time-invariant Markov process ($P_{X_{t+1}|X_t}$ does not depend on $t$), the 
channel is time-invariant (transition matrix $P_t$ does not depend on $t$) 
and the distortion metric $\rho_t(\cdot)$ is time invariant. For such a 
system, the functions $\nu_t(\cdot)$, $\psi_t(\cdot)$, the linear transforms 
$Q_t(\cdot)$, $\widehat Q_t(\cdot)$ and the distortion 
$\tilde{\rho}_t(\cdot)$ defined in Theorem~\ref{thm:info-state} are 
time-invariant, so we can drop the subscripts $t$ and simply refer them as 
$\nu(\cdot)$, $\psi(\cdot)$, $Q(\cdot)$, $\widehat Q(\cdot)$ and 
$\tilde{\rho}(\cdot)$ respectively. So, we obtain an equivalent of 
Theorem~\ref{thm:info-state} making the corresponding changes.
Thus, Problem~\ref{prob:P3} reduces to a time-homogeneous problem --- one in 
which state space, action space, system update equation and instantaneous 
cost do not depend on time. Hence the optimality equations of 
Theorem~\ref{thm:info-state} can be written in a more compact manner We can 
define the following:
\begin{definition}\label{def:operators}
  Let $\V$ (resp. $\widehat\V$) be the family of functions from $\Pi$ (resp.  
  $\Phi$) to $\mathds R^+$. Define operators $W(c)$ (resp. $\widehat W(l)$) 
  from $\widehat\V$ to $\V$ (resp. $\V$ to $\widehat\V$) as follows:
  \begin{align}
    \big(W(c)\widehat V\big)(\pi) &= \tilde\rho\big(Q(c)\pi\big) + \widehat 
    V\big(Q(c)\pi\big), \\
    \big(\widehat W(l)V\big)(\varphi) &= V\big(\widehat Q(l)\varphi\big).
  \end{align}
  Further define transformations $\W$ (resp. $\widehat\W$) from $\widehat\V$ 
  to $\V$ (resp. $\V$ to $\widehat\V$) as follows:
  \begin{align}
    \big( \W\widehat V \big)(\pi) &= \inf_{c \in \sC_S} \big( W(c)\widehat V 
    \big)(\pi), \\
    \big( \widehat\W V \big)(\varphi) &= \inf_{l \in \sL}\big( \widehat W(l)V 
    \big)(\varphi).
  \end{align}
\end{definition}
\begin{theorem}\label{thm:homogenous-DP}
  For the time-homogeneous case, the value functions $V_t$ and $\widehat V_t$ 
  of Theorem~\ref{thm:DP} evolve in a time-homogeneous manner as follows:
  \begin{alignat}{2}
    \widehat V_t &= \widehat \W V_{t+1},&\qquad t &=1,\dots,T-1, \\
    V_t &= \W \widehat V_t, &\qquad t &= 1,\dots,T,
  \end{alignat}
  with the terminal condition given by
  \begin{equation}
    \widehat V_T \equiv 0.
  \end{equation}
  The arguments minimizing $V_t$ and $\widehat V_t$ at each stage determine 
  the decision rules $c_t$ and $l_t$.
\end{theorem}
\begin{proof}
  This follows immediately from Theorem~\ref{thm:DP}.
\end{proof}

\vspace{-1em}
\section{Time Homogeneous System --- Infinite Horizon Case}
\label{sec:infinite}
\vspace{-1em}
We consider a time-homogeneous system as in Section~\ref{sec:finite}. However,
instead of a finite horizon $T$, we consider the infinite horizon case with
performance of a design determined by
\begin{equation}
  \J(c,g,l) = \E{\sum_{t=1}^{\infty}\beta^{t-1}\rho(X_t,\widehat X_t) | c, g, 
  l},
  \label{eq:infinite-performance}
\end{equation}
where $\beta \in (0,1)$ is called the discount factor. With a slight 
modification of the proof of~\cite[Section~2.4]{Teneketzis:2004} one can show
that the structural results of Section~\ref{sec:structural} are also valid in
this case. Further, Theorem~\ref{thm:info-state} (with the changes mentioned 
in previous section) holds for the infinite-horizon case also. 

\begin{definition}
  Define operators $W$, $\widehat W$ and transforms $\W$, $\widehat\W$ as in 
  Definition~\ref{def:operators}, with one change --- modify the definition 
  of $W$ to take the discounting into account as follows:
  \begin{equation}
    \big( W(c)\widehat V \big)(\pi) = \tilde\rho\big( Q(c)\pi \big) + 
    \beta\widehat V\big( Q(c)\pi \big).
  \end{equation}
\end{definition}
\begin{theorem}
  \label{thm:VF-evolution}
  For the infinite horizon time-homogeneous system with the performance 
  criterion of \eqref{eq:infinite-performance}, the evolution of value 
  function is governed by the following set of equations
  \begin{align}
    V_t &= \W \widehat V_{t}, \\
    \widehat V_t &= \widehat\W V_{t+1}.
  \end{align}
  The arguments that minimize $\widehat V_{t}$ and $V_{t+1}$ at each stage 
  determine the decision rules $c_t$ and $l_t$.
\end{theorem}
\begin{proof}
  This is the solution of the time-homogeneous problem formulated by 
  considering a time-homogeneous version of Problem~\ref{prob:P3} with the 
  optimization criteria being minimizing 
  $\E{\sum_{t=1}^{\infty}\beta^{t-1}\tilde\rho(\varphi_t) | c, g^*(c,l), l}$.
\end{proof}

\begin{definition}
  A design $(c,l)$, $c \defined (c_1,c_2,\dots)$, $l \defined 
  (l_1,l_2,\dots)$ is called stationary (or time-invariant) if $c_1 \equiv 
  c_2 \equiv \dots \equiv c$, $l_1 \equiv l_2 \equiv \dots \equiv l$.
\end{definition}

\begin{theorem}
  \label{thm:stationary_policy}
  For the time homogeneous case with the performance measure 
  of~\eqref{eq:infinite-performance}, if the distortion measure $\rho(\cdot)$ 
  is bounded and discount factor $\beta < 1$, then stationary designs are 
  $\varepsilon$-optimal, that is, for any design $(c',l')$ and any 
  $\varepsilon > 0$, there exists a stationary design $(c^\infty,l^\infty)$ 
  such that
  \begin{equation}
    \J(c^\infty,l^\infty) = V(\pi_1) \leq \J(c',l') + \varepsilon,
  \end{equation}
  where $V$ is the \emph{unique} fixed point of
  \begin{equation}
    V = (\W\circ\widehat \W) V,
  \end{equation}
  and $c$ and $l$ are the corresponding $\arg\min$ and $c^\infty = 
  (c,c,\dots)$, $l^\infty = (l,l,\dots)$.
\end{theorem}
\begin{proof} See~\cite{MahajanTeneketzis:2005}. \end{proof}
We have shown that a unique stationary $\varepsilon$-optimal policy exists. Thus, 
for the infinite horizon problem, without loss of optimality, we can restrict 
attention to stationary policies. This simplifies the implementation of an 
optimal policy.

\vspace{-1em}
\section{Discussion}\label{sec:discussion}
\vspace{-1em}
It was  shown in~\cite{Witsenhausen:1973} that all sequential problems can be 
transformed to a \emph{standard form} by moving all the uncertainty to the
first stage and at each stage augmenting the state variable to carry all the
information needed to determine the cost. Further an optimal policy for a
problem in standard form can be obtained by solving a deterministic
optimization problem. We believe that our methodology has a similar spirit as
Witsenhausen's standard form. We have a decentralized optimization problem
that is sequential and an optimal design is obtained by the solution of a
deterministic optimization problem. In our solution the state space is not 
increasing with time and allows us to use our approach to infinite horizon 
problems while the standard form is applicable only to finite horizon 
problems.  The structural results of~\cite{Teneketzis:2004} are critical to 
our approach as they allow us to obtain an information state whose 
dimensionality does not change with time.

\vspace{-1em}
\section{Conclusion}\label{sec:conclusion}
\vspace{-1em}
We have developed a methodology for the determination of jointly optimal 
real-time encoding, decoding and memory update strategies for point-to-point 
communication system. This methodology has been extended to $k$-th oder 
Markov sources, distortion metric accepting a finite delay of $\delta$ units 
and channels with memory (see~\cite{MahajanTeneketzis:2005} for details). We 
believe that the same methodology can be used for the determination of 
jointly optimal real-time encoding, decoding and memory update strategies for 
more complex communication systems.

\vspace{-1em}
\bibliographystyle{IEEEtranS}
\bibliography{IEEEfull,../collection}

\begin{thebibliography}{10}
\providecommand{\url}[1]{#1}
\csname url@rmstyle\endcsname
\providecommand{\newblock}{\relax}
\providecommand{\bibinfo}[2]{#2}
\providecommand\BIBentrySTDinterwordspacing{\spaceskip=0pt\relax}
\providecommand\BIBentryALTinterwordstretchfactor{4}
\providecommand\BIBentryALTinterwordspacing{\spaceskip=\fontdimen2\font plus
\BIBentryALTinterwordstretchfactor\fontdimen3\font minus
  \fontdimen4\font\relax}
\providecommand\BIBforeignlanguage[2]{{%
\expandafter\ifx\csname l@#1\endcsname\relax
\typeout{** WARNING: IEEEtran.bst: No hyphenation pattern has been}%
\typeout{** loaded for the language `#1'. Using the pattern for}%
\typeout{** the default language instead.}%
\else
\language=\csname l@#1\endcsname
\fi
#2}}

\bibitem{BorkarMitterTatikonda:2001}
V.~Borkar, S.~Mitter, and S.~Tatikonda, ``Optimal sequential vector
  quantization of {Markov} sources,'' \emph{{SIAM} Journal of Optimal Control},
  vol.~40, no.~1, pp. 135--148, Jan 2001.

\bibitem{LipsterShiryayen:II}
R.~Lipster and A.~Shiryayev, \emph{Statistics of Random Processes,
  Vol.~II:Applications}.\hskip 1em plus 0.5em minus 0.4em\relax
  Springer-Verlag, 1977.

\bibitem{MahajanTeneketzis:2005}
A.~Mahajan and D.~Teneketzis, ``On jointly optimal encoding, decoding and
  memory update for noisy real-time communication systems,'' Department of
  {EECS}, {U}niversity of {M}ichigan, Ann Arbor, MI--48109-2122, Control Group
  Report CGR-05-07, Oct. 2005, to be submitted to IEEE Trans. on Information
  Theory.

\bibitem{Teneketzis:2004}
D.~Teneketzis, ``On the structure of optimal real-time encoders and decoders in
  noisy communiation,'' submitted for publication in IEEE Trans. on Information
  Theory.

\bibitem{WalrandVaraiya:1983}
J.~C. Walrand and P.~Varaiya, ``Optimal causal coding---decoding problems,''
  \emph{{IEEE} Transactions on Information Theory}, vol.~29, no.~6, pp.
  814--820, Nov. 1983.

\bibitem{Whittle:1982}
P.~Whittle, \emph{Optimization Over Time}, ser. Wiley series in Probability and
  Mathematical Statistics.\hskip 1em plus 0.5em minus 0.4em\relax John Wiley
  and Sons, 1982, vol.~1.

\bibitem{Witsenhausen:1971}
H.~S. Witsenhausen, ``Separation of estimation and control for discrete time
  systems,'' \emph{Proceedings of the {IEEE}}, vol.~59, no.~11, pp. 1557--1566,
  Nov. 1971.

\bibitem{Witsenhausen:1973}
------, ``A standard form for sequential stochastic control.''
  \emph{Mathematical Systems Theory}, vol.~7, no.~1, pp. 5--11, 1973.

\bibitem{Witsenhausen:1975c}
------, ``The instrinsic model for stochastic control: Some open problems,'' in
  \emph{Lecture Notes in Economics and Mathematical Systems, 107}.\hskip 1em
  plus 0.5em minus 0.4em\relax Springer Verlag, 1975, pp. 322--335.

\bibitem{Witsenhausen:1976}
------, ``Some remark on the concept of state,'' in \emph{Directions in
  Large-Scale Systems}, Y.~Ho and S.~Mitter, Eds.\hskip 1em plus 0.5em minus
  0.4em\relax Plenum, 1976, pp. 69--75.

\bibitem{Witsenhausen:1978}
------, ``On the structure of real-time source coders,'' \emph{Bell System
  Technical Journal}, vol.~58, no.~6, pp. 1437--1451, July-August 1978.

\end{thebibliography}
\end{document}